\documentclass[12pt]{article}\usepackage{qGibbs}
\input{MyQcircuit.sty}
\begin{document}

\title{Quantum Gibbs Sampling \\
Using Szegedy Operators}

\author{Robert R. Tucci\\
        P.O. Box 226\\
        Bedford,  MA   01730\\
        tucci@ar-tiste.com}

\date{ \today}

\maketitle

\vskip2cm
\section*{Abstract}
We present an algorithm for
doing Gibbs sampling
on a quantum computer.
The algorithm combines
 phase estimation for a Szegedy operator,
 and Grover's algorithm.
For any $\epsilon>0$,
the algorithm will
sample a probability distribution
in ${\cal O}(\frac{1}{\sqrt{\delta}})$
steps with precision ${\cal O}(\epsilon)$.
Here $\delta$ is the distance
between the two largest
eigenvalue magnitudes
of the transition matrix
of the Gibbs Markov chain used in the algorithm.
It takes ${\cal O}(\frac{1}{\delta})$
steps to achieve
the same precision
if one does Gibbs sampling
on a classical computer.
\newpage

\section{Introduction}

In Ref.\cite{szegedy},
Szegedy proposed a quantum
walk operator
for
each classical Markov chain.
In Ref.\cite{somma},
Somma et al.
proposed a method for doing
simulated annealing
on a quantum computer.
In Ref.\cite{wocjan1},
Wocjan et al.
showed how to improve
the Somma et al. algorithm.
The algorithms of Somma et al.
and Wocjan et al. both
use Szegedy
operators.
In Ref.\cite{TucQusann}, I
presented computer
programs called QuSAnn
and Multiplexor Expander
that implement
ideas of Refs.\cite{somma} and \cite{wocjan1},
and also some of my own
ideas about quantum multiplexors.

In Ref.\cite{TucMetHas},
I described one particular
algorithm for doing Gibbs
and
Metropolis-Hastings sampling
of a classical Bayesian network
(i.e., a probability distribution)
on a quantum computer.
In this paper, I describe a
different algorithm for
doing Gibbs sampling
on a quantum computer.
Unlike my first algorithm,
this one uses Szegedy
operators.
For any $\epsilon>0$,
this new algorithm will
sample a Bayesian network
in ${\cal O}(\frac{1}{\sqrt{\delta}})$
steps with precision ${\cal O}(\epsilon)$.
Here $\delta$ is the distance
between the two largest
eigenvalue magnitudes
of the transition matrix
of the Gibbs Markov chain used in the algorithm.
It takes ${\cal O}(\frac{1}{\delta})$
steps to achieve
the same precision
if one does Gibbs sampling
on a classical computer.

This paper assumes that its
reader has read the section
entitled ``Notation
and Preliminaries" in Ref.\cite{TucMetHas}.
The reader should refer
to Refs.\cite{TucMetHas,TucQusann}
for clarification when any
notation of this paper
eludes him.

\section{Dual Gibbs Markov Chains}
In this section, we will
discuss dual ``Gibbs" Markov chains with
transition matrices $M_1$ and $M_2$, respectively.
These two transition matrices are both
defined in terms of a single
classical Bayesian network $\rvx$.

\subsection{Definitions of $M_1$ and $M_2$}

Consider a
classical Bayesian net
with $N_{nds}$ nodes, labeled
$\rvx_1,\rvx_2,\ldots,\rvx_{N_{nds}}$
where $\rvx_j\in S_{\rvx_j}$
for each $j$. (As usual in my papers, I
indicate random variables
by underlining them.)
Let $\rvx = (\rvx_1,\rvx_2,\ldots,\rvx_{N_{nds}})$.
Let $\rvx$ assume values in a set
$S_\rvx$ which has $N_S=2^\nb$ elements.

Let

\beq
\pi(x) = P(\rvx=x)
\;
\eeq
for all $x\in S_\rvx$.

For $N_{nds}=3$ and $x,y\in S_\rvx$,
let

\beq
M_1(y|x)=
P_{\rvx_1|\rvx_2\rvx_3}(y_1|x_2,x_3)
P_{\rvx_2|\rvx_3\rvx_1}(y_2|x_3,y_1)
P_{\rvx_3|\rvx_1\rvx_2}(y_3|y_1,y_2)
\;,
\eeq
and

\beq
M_2(y|x)=
P_{\rvx_1|\rvx_2\rvx_3}(y_1|y_2,y_3)
P_{\rvx_2|\rvx_3\rvx_1}(y_2|y_3,x_1)
P_{\rvx_3|\rvx_1\rvx_2}(y_3|x_1,x_2)
\;.
\eeq
($M_2(y|x)$ can
be obtained
by swapping $x_i$ and $y_i$
in the {\it conditioned} arguments of
$M_1(y|x)$.)
Note that $\sum_y M_1(y|x)=1$
and $\sum_y M_2(y|x)=1$.
Define $M_1$ and $M_2$
for arbitrary $N_{nds}$
using the same pattern.
$M_1$ and $M_2$
are transition
matrices
of the type typical for Gibbs sampling.
(See
Ref.\cite{TucMetHas} for an introduction
to Gibbs sampling and the more general
Metropolis-Hastings sampling).

You can check that  $\pi()$
is not a detailed balance of
either $M_1$ nor
$M_2$ separately. However,
the following
property is true.
We will refer to this property
by saying that
$\pi()$
is a detailed balance of
the pair $(M_1,M_2)$.

\begin{claim}
\beq
M_1(y|x)\pi(x)=M_2(x|y)\pi(y)
\;
\label{eq-pair-detBal}
\eeq
for all $x,y\in S_\rvx$.
\end{claim}
\proof

Let
$P(x_j,x_k,\ldots)$
stand for $P(\rvx_j=x_j, \rvx_k=x_k,\ldots)$.
Assume $N_{nds}=3$ to begin with. One has

\beqa
\frac{M_1(y|x)}{M_2(x|y)}&=&
\frac
{
P(y_1|x_2,x_3)
P(y_2|x_3,y_1)
P(y_3|y_1,y_2)
}{
P(x_1|x_2,x_3)
P(x_2|x_3,y_1)
P(x_3|y_1,y_2)
}
\\
&=&
\frac
{
P(y_1,x_2,x_3)
P(y_2,x_3,y_1)
P(y_3,y_1,y_2)
}{
P(x_1,x_2,x_3)
P(x_2,x_3,y_1)
P(x_3,y_1,y_2)
}
\\
&=&
\frac
{
P(y_1,x_2,x_3)
P(y_1,y_2,x_3)
P(y)
}{
P(x)
P(y_1,x_2,x_3)
P(y_1,y_2,x_3)
}
\\
&=&
\frac{P(y)}{P(x)}
\;.
\eeqa
A proof for an arbitrary number $N_{nds}$
of nodes follows the same pattern.
\qed

\subsection{Eigenvalues of $M_1$,
$M_2$ and $M_{hyb}$}

Let

\beq
\Lam_j(y|x)=\sqrt{M_j(y|x)}
\;,
\eeq
for $j=1,2$ and $x,y\in S_\rvx$.
It's convenient to
define
a hybrid function of $M_1$ and $M_2$,
as follows:

\beq
M_{hyb}(y|x) = \Lam_2(x|y)\Lam_1(y|x)
\;
\eeq
for $x,y\in S_\rvx$.
(Note that unlike $M_1(y|x)$ and $M_2(y|x)$,
$M_{hyb}(y|x)$ is not a probability
function in $y$, its first
argument.)

Define the quantum states

\beq
\ket{(\pi)^\eta}=
\sum_x [\pi(x)]^\eta\ket{x}
\;
\eeq
for $\eta=\frac{1}{2}, 1$. (Note that only the
$\eta=\frac{1}{2}$
state is normalized in the sense
of quantum mechanics.)

\begin{claim}

\beq
M_j\ket{\pi}=\ket{\pi}
\;\;\mbox{for }j=1,2
\;,
\eeq
and

\beq
M_{hyb}\ket{\sqrt{\pi}}=\ket{\sqrt{\pi}}
\;.
\eeq
Also, $M_1$, $M_2$ and $M_{hyb}$ have the same
eigenvalues.
\end{claim}
\proof

Taking the square root
of both sides of the pair detailed balance
statement Eq.(\ref{eq-pair-detBal}), we get

\beq
\Lam_1(y|x)\sqrt{\pi(x)}=
\Lam_2(x|y)\sqrt{\pi(y)}
\;.
\eeq
Therefore,

\beqa
M_{hyb}(y|x)=\Lam_2(x|y)
\frac{1}{\sqrt{\pi(x)}}
\Lam_2(x|y)\sqrt{\pi(y)}
&=&
\frac{1}{\sqrt{\pi(x)}}
M_2(x|y)
\sqrt{\pi(y)}
\;.
\eeqa
Hence,

\beq
\sum_x M_1(y|x)\pi(x)=
\sum_x M_2(x|y)\pi(y)=\pi(y)
\;,
\eeq

\beq
\sum_y M_2(x|y)\pi(y)=
\sum_y M_1(y|x)\pi(x)=\pi(x)
\;,
\eeq
and

\beq
\sum_x M_{hyb}(y|x)\sqrt{\pi(x)}
=
\sum_x
\frac{1}{\sqrt{\pi(x)}}M_2(x|y)\sqrt{\pi(y)}
\sqrt{\pi(x)}
=\sqrt{\pi(y)}
\;.
\eeq

Order the elements of
the finite set $S_\rvx$
in some preferred way.
Use this preferred order
to
represent $M_1$, $M_2$ and $M_{hyb}$
as matrices.
Define a diagonal matrix $D$
whose diagonal entries are the
numbers $\pi(x)$ for each $x\in S_\rvx$,
with the $x$ ordered in the
preferred order:

\beq
D= diag[\left(\pi(x)\right)_{\forall x}]
\;.
\eeq
Since

\beq
M_2^T = D^{-1}M_1 D\;\;,
M_{hyb}^T = D^{-\frac{1}{2}}M_2 D^{\frac{1}{2}}
\;,
\eeq
it follows that

\beq
\det(M_1 - \lambda)=
\det(M_2 - \lambda)=
\det(M_{hyb} - \lambda)
\;
\eeq
for any $\lambda\in \CC$.
\qed

Let the eigenvalues\footnote{There must
be a single eigenvalue 1 and all others
must have a magnitude strictly smaller
than one because of the Frobenius-Perron Theorem.
The eigenvalues may be complex.}
 of $M_{hyb}$ (and also of $M_1$ and $M_2$)
be $m_0, m_1, \ldots m_{\ns-1}\in\CC$ with
$m_0=1> |m_1|\geq
|m_2|\ldots\geq|m_{\ns-1}|$.
Define $\ket{m_j}$ to be the corresponding
eigenvectors of $M_{hyb}$ (but not necessarily
of $M_1$ and $M_2$). Thus

\beq
M_{hyb}\ket{m_j} = m_j\ket{m_j}
\;,
\eeq
for $j=0,1,\ldots,\ns-1$.
In particular, $\ket{m_0}=\ket{\sqrt{\pi}}$.

For each $j$,
define $\varphi_j\in[0,\frac{\pi}{2}]$
and $\eta_j\in[0,2\pi)$ so that
$m_j = e^{i\eta_j}\cos\varphi_j $.
(Thus, $\cos\varphi_j\geq 0$).
Note that $m_0=1$ so $\varphi_0=0$.
The $M_1$ eigenvalue gap $\delta$ is defined as
$\delta=1-|m_1|$. $\delta\approx\frac{\varphi_1^2}{2}$
when $\varphi_1$ is small.

\section{Q-Embeddings $U_1$ and $U_2$}
In this section, we will
define a ``q-embedding" $U_j$
of $M_j$, for $j=1,2$.
(For more information about
q-embeddings, see Ref.\cite{TucMetHas}.)

For simplicity,
we begin this section by
considering a Bayesian net
with only 3 nodes $\rvx_1,\rvx_2,\rvx_3$,
and such that each of
these nodes is binary (i.e.,
$S_{\rvx_j}=Bool$ for $j=1,2,3$).
At the end of this section, we will
show how to
remove these
restrictions
and make our treatment valid
for general
Bayesian networks.

\begin{figure}[h]
    \begin{center}
    \epsfig{file=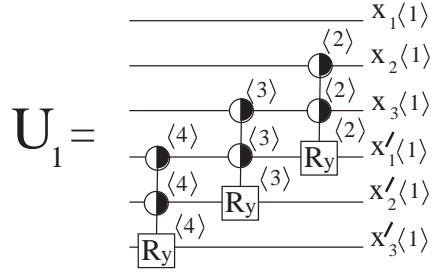, height=1.5in}
    \caption{Unitary matrix $U_1$
    expressed as a product of quantum multiplexors.}
    \label{fig-U1}
    \end{center}
\end{figure}

Using the same language
as Ref.\cite{TucMetHas},
consider a unitary matrix $U_1$
of the form shown in Fig.\ref{fig-U1},
with its multiplexor gates defined as follows.
Let $\rvx_j\av{k}\in Bool$
and $\rvx'_j\av{k}\in Bool$ for any
$j,k$.
 $U_1$ has
3 analogous gates (a.k.a. nodes)
labeled
$(\rvx'_{1}\av{2},
\rvx_{3}\av{2},
\rvx_{2}\av{2})$,
$(\rvx'_{2}\av{3},
\rvx'_{1}\av{3},
\rvx_{3}\av{3})$,
and
$(\rvx'_{3}\av{4},
\rvx'_{2}\av{4},
\rvx'_{1}\av{4})$.
Consider
the first of these for definiteness.
Let the probability amplitude
$A(x'_1\av{2},x_3\av{2},x_2\av{2}|
x'_1\av{1},x_3\av{1},x_2\av{1})$
of node $(\rvx'_1\av{2},\rvx_3\av{2},\rvx_2\av{2})$
satisfy the constraint

\beqa
\lefteqn{A(x'_1\av{2},x_3\av{2},x_2\av{2}|
x'_1\av{1}=0,x_3\av{1},x_2\av{1})}\\
&=&
\sqrt{P_{\rvx_1|\rvx_3,\rvx_2}(
x'_1\av{2}|x_3\av{2},x_2\av{2})}
\delta_{x_2\av{2}}^{x_2\av{1}}
\delta_{x_3\av{2}}^{x_3\av{1}}
\;.
\label{eq-A-constraint}
\eeqa
If we indicate
non-zero entries by a plus sign,

\beqa
A &=&
\begin{tabular}{r|r|r|r|r|r|}
     & {\tiny 000}& {\tiny 001} &{\tiny 010}& {\tiny 011} & $\cdots$ \\
\hline
{\tiny $(x'_1,x_3,x_2)$= 000} &+& & &&$\cdots$ \\
{\tiny 001} &  &+&&&$\cdots$\\
{\tiny 010} &  &&+&&$\cdots$\\
{\tiny 011} &  &&&+&$\cdots$\\
{\tiny 100} &  +&&&&$\cdots$\\
{\tiny 101} &  &+&&&$\cdots$\\
{\tiny 110} &  &&+&&$\cdots$\\
{\tiny 111} &  &&&+&$\cdots$\\
\end{tabular}
\\
&\rarrow&
\sum_{\vecb\in Bool^2}
e^{i\theta_\vecb\sigy}\otimes P_\vecb
=
\begin{array}{c}
\Qcircuit @C=1em @R=1em @!R{
&\muxorgate
&\qw
\\
&\muxorgate
&\qw
\\
&\emptygate\qwx[-2]
&\qw
}
\end{array}
\;,
\label{eq-reduced-a}
\eeqa
for some $\theta_\vecb\in \RR$.
Here the right pointing
arrow means that the expression at the
origin of the arrow can
be extended to the expression
at the target of the arrow.

From the above definition of $U_1$,
it follows that,
for $x,x',y,y'\in Bool^3$,

\beqa
\begin{array}{c}
\bra{y}
\\
\bra{y'}
\end{array}
U_1
\begin{array}{l}
\ket{x}
\\
\ket{0}^{\otimes 3}
\end{array}
&=&
\left\{
\;\;\;\;\;
\begin{array}{c}
\Qcircuit @C=1em @R=1em @!R{
\bra{y_1}\;\;\;\;\;
&\qw
&\qw
&\qw
&\qw\;\;\;\;\;\ket{x_1}
\\
\bra{y_2}\;\;\;\;\;
&\qw
&\qw
&\muxorgate
&\qw\;\;\;\;\;\ket{x_2}
\\
\bra{y_3}\;\;\;\;\;
&\qw
&\muxorgate
&\muxorgate
&\qw\;\;\;\;\;\ket{x_3}
\\
\bra{y'_1}\;\;\;\;\;
&\muxorgate
&\muxorgate
&\emptygate\qwx[-2]
&\qw\;\;\;\;\;\ket{0}
\\
\bra{y'_2}\;\;\;\;\;
&\muxorgate
&\emptygate\qwx[-2]
&\qw
&\qw\;\;\;\;\;\ket{0}
\\
\bra{y'_3}\;\;\;\;\;
&\emptygate\qwx[-2]
&\qw
&\qw
&\qw\;\;\;\;\;\ket{0}
}
\end{array}
\right.
\\
&=&
\Lam_1(y'|x)\delta(y,x)
\;.
\eeqa
Hence,

\beq
\begin{array}{c}
\Qcircuit @C=1em @R=2em @!R{
&\multigate{1}{\mbox{$U_1$}}
&\qw
&\ket{x}
\\
&\ghost{1}{\mbox{$U_1$}}
&\qw
&\ket{0}^{\otimes 3}
}
\end{array}
\;\;
=
\;\;
\begin{array}{c}
\Qcircuit @C=1em @R=1em @!R{
&\qw
&\qw
&\ket{x}
\\
&\gate{\Lam_1}
&\qw
&\ket{x}
}
\end{array}
\mbox{\;\;\;\;or    }
U_1\ket{0}^{\otimes 3}\ket{x}=
(\Lam_1\ket{x})\ket{x}
\;.
\eeq

\begin{figure}[h]
    \begin{center}
    \epsfig{file=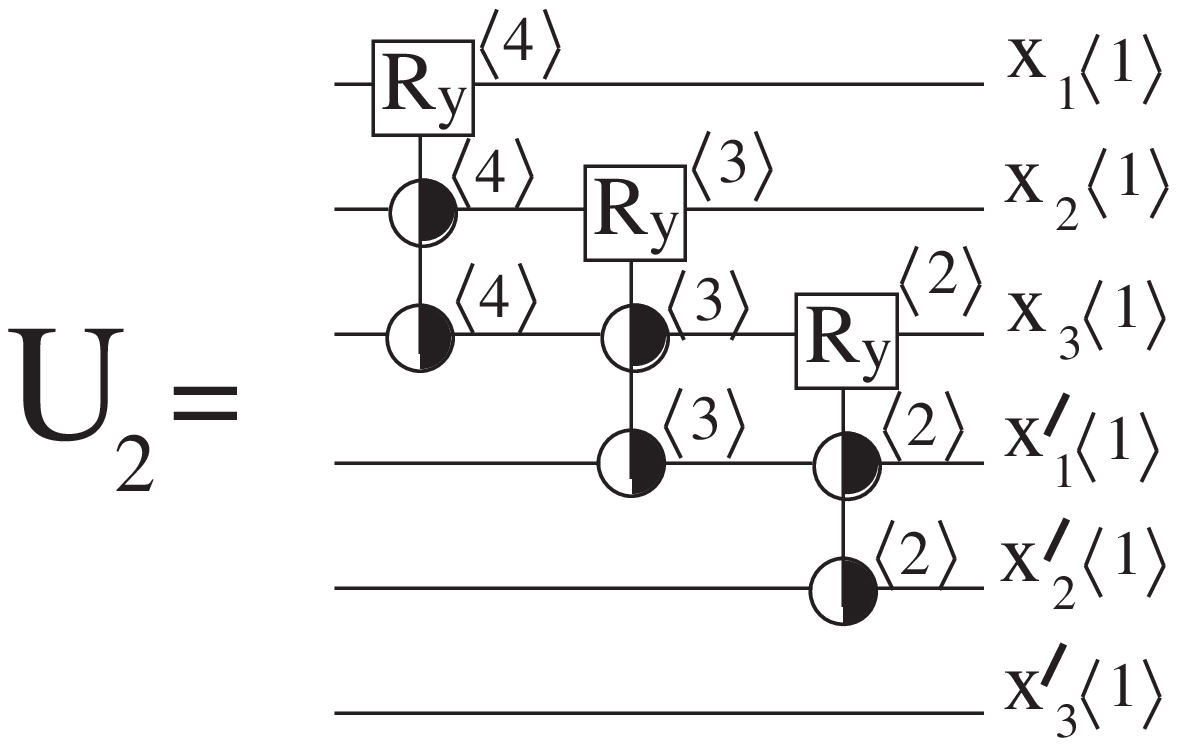, height=1.5in}
    \caption{Unitary matrix $U_2$
    expressed as a product of quantum multiplexors.}
    \label{fig-U2}
    \end{center}
\end{figure}

Besides $U_1$,
it is convenient to
consider another unitary matrix
called $U_2$.
We define $U_2$
to be of the form of
Fig.\ref{fig-U2}, where the multiplexors
are defined in such a way that
$U_2$ satisfies,
for all $x,x',y,y'\in Bool^3$,

\beqa
\begin{array}{c}
\bra{y}
\\
\bra{y'}
\end{array}
U_2
\begin{array}{l}
\ket{0}^{\otimes 3}
\\
\ket{x'}
\end{array}
&=&
\left\{
\;\;\;\;\;
\begin{array}{c}
\Qcircuit @C=1em @R=1em @!R{
\bra{y_1}\;\;\;\;\;
&\emptygate\qwx[2]
&\qw
&\qw
&\qw\;\;\;\;\;\ket{0}
\\
\bra{y_2}\;\;\;\;\;
&\muxorgate
&\emptygate\qwx[2]
&\qw
&\qw\;\;\;\;\;\ket{0}
\\
\bra{y_3}\;\;\;\;\;
&\muxorgate
&\muxorgate
&\emptygate\qwx[2]
&\qw\;\;\;\;\;\ket{0}
\\
\bra{y'_1}\;\;\;\;\;
&\qw
&\muxorgate
&\muxorgate
&\qw\;\;\;\;\;\ket{x_1'}
\\
\bra{y'_2}\;\;\;\;\;
&\qw
&\qw
&\muxorgate
&\qw\;\;\;\;\;\ket{x_2'}
\\
\bra{y'_3}\;\;\;\;\;
&\qw
&\qw
&\qw
&\qw\;\;\;\;\;\ket{x_3'}
}
\end{array}
\right.
\\
&=&
\Lam_2(y|x')\delta(y',x')
\;.
\eeqa
Hence

\beq
\begin{array}{c}
\Qcircuit @C=1em @R=2em @!R{
&\multigate{1}{\mbox{$U_2$}}
&\qw
&\ket{0}^{\otimes 3}
\\
&\ghost{1}{\mbox{$U_2$}}
&\qw
&\ket{x'}
}
\end{array}
\;\;
=
\;\;
\begin{array}{c}
\Qcircuit @C=1em @R=1em @!R{
&\gate{\Lam_2}
&\qw
&\ket{x'}
\\
&\qw
&\qw
&\ket{x'}
}
\end{array}
\mbox{\;\;\;\;or    }
U_2\ket{x'}\ket{0}^{\otimes 3}=
\ket{x'}(\Lam_2\ket{x'})
\;.
\eeq

$U_j$ is called the q-embedding
of $M_j$ for $j=1,2$.

\begin{figure}[h]
    \begin{center}
    \epsfig{file=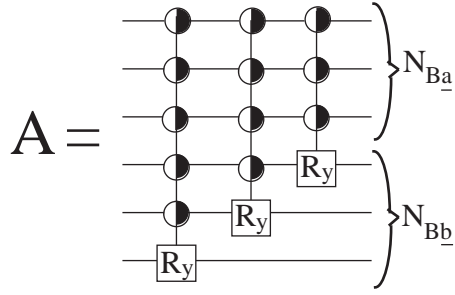, height=1.5in}
    \caption{The unitary matrix
    $A$ is a quantum
    embedding of a
    probability matrix $P(b|a)$,
    where $a$ has $\nba$ bits
    and $b$ has $\nbb$ bits.
    }
    \label{fig-multi-qubit-nodes}
    \end{center}
\end{figure}

So far we have considered
the q-embeddings $U_1$
and $U_2$ for
the case of a classical
Bayesian network $\rvx$ with
3 binary nodes.
What if $\rvx$ has $N_{nds}$ nodes
and some of those
nodes have more than
2 states? In that case, we must
use several qubits (horizontal lines) for
each node $\rvx_i$
(and an equal number of
qubits for the dual node $\rvx'_i$)
in Figs.\ref{fig-U1} and \ref{fig-U2}.
More specifically, suppose
$P(x_1|x_2,x_3,\ldots,x_{N_{nds}})$
equals
$P(b|a)$ where $a\in Bool^{\nba}$
and $b\in Bool^{\nbb}$.
For the number of bits $\nba$,
define the number of states
$\nsa = 2^{\nba}$. Likewise, let
$\nsb = 2^{\nbb}$.
The constraint Eq.(\ref{eq-A-constraint})
generalizes to

\beq
A(b,\tilde{a}| \tilde{b}=0, a)=
\sqrt{P(b|a)}\delta_a^{\tilde{a}}
\;,
\label{eq-amp-tensor-form}
\eeq
where
$a,\tilde{a}\in Bool^{\nba}$
and $b,\tilde{b}\in Bool^{\nbb}$.
Eq.(\ref{eq-amp-tensor-form})
can be expressed in matrix form as follows:

\beq
[A(b,\tilde{a}|\tilde{b}=0,a)]=
\begin{array}{c|c}
&(\tilde{b}=0,a)\rarrow\\
\hline
(b,\tilde{a})&D^{0,0}\\
\downarrow&D^{1,0}\\
&\cdots \\
&D^{\nsb-1,0}
\end{array}
\;,
\label{eq-amp-mat-form-one-col}
\eeq
where,
for all $b\in Bool^{\nbb}$,
 $D^{b,0}\in \RR^{\nsa X \nsa}$
 are diagonal matrices
 with entries

 \beq
 (D^{b,0})_{a,\tilde{a}}=\sqrt{P(b|a)}\delta_a^{\tilde{a}}
 \;.
 \eeq
By adding
 more columns to the matrix of
 Eq.(\ref{eq-amp-mat-form-one-col}),
 one can
extended
it (see section entitled
``Q-Embeddings" in Ref.\cite{TucMetHas}) to a
square matrix which can be expressed
in terms of multiplexors as in
Fig.\ref{fig-multi-qubit-nodes}.

The Markov Blanket
$MB(i)$ for a node $\rvx_i$
of the classical Bayesian network $\rvx$ satisfies
(see section entitled
``Notation and Preliminaries"
in  Ref.\cite{TucMetHas})

\beq
P(x_i|x_\noti)=P(x_i|x_{MB(i)})
\;.
\eeq
If the set $MB(i)$
is strictly smaller
than the set $\noti$,
this property can be used to
reduce the number of controls
for the multiplexor
in $U_1$ and $U_2$
corresponding
to $P(x_i|x_\noti)$.

Given the two q-embeddings $U_1$
and $U_2$ for a Bayesian network $\rvx$,
we can define a unitary matrix $U$
as follows

\beq
U= U_2^\dagger U_1
\;.
\eeq
Matrix $U$ has the
following highly desirable property:

\begin{claim}
For any $j,k\in\{0,1,\dots,N_S-1\}$,

\beq
\begin{array}{r}
\bra{0}
\\
\bra{m_j}
\end{array}
U
\begin{array}{l}
\ket{m_k}
\\
\ket{0}
\end{array}
=
m_j \delta_j^k
\;.
\eeq
\end{claim}
\proof
\beqa
\begin{array}{r}
\bra{0}
\\
\bra{m_j}
\end{array}
U^\dagger_2 U_1
\begin{array}{l}
\ket{m_k}
\\
\ket{0}
\end{array}
&=&
\sum_{y,x}
\begin{array}{r}
\;
\\
\av{m_j|y}
\end{array}
\left[
\begin{array}{l}
\bra{y}\Lam_2^T
\\
\bra{y}
\end{array}
\right]\left[
\begin{array}{r}
\ket{x}
\\
\Lam_1\ket{x}
\end{array}
\right]
\begin{array}{l}
\av{x|m_k}
\\
\;
\end{array}
\\
&=&
\sum_{y,x}
\av{m_j|y}\Lam_2^T(y|x)\Lam_1(y|x)\av{x|m_k}
\\
&=&
\av{m_j|M_{hyb}|m_k}=
m_j \delta_j^k
\;.
\eeqa
\qed

\section{Szegedy Quantum Walk Operator $W$}
In this section,
we will define a
special type of Szegedy
quantum walk operator $W$
corresponding to a Bayesian net $\rvx$.
We will then find the eigenvalues
of $W$.

\subsection{Definition of $W$}

As in Ref.\cite{TucQusann}, define
the projection operator $\hat{\pi}$
and its dual projection operator $\check{\pi}$
by

\beq
\hat{\pi}=
\begin{array}{c}
\ket{0}\bra{0}
\\
\mbox{---}
\end{array}
\;,\;\;
\check{\pi}=
\swap \hat{\pi} \swap=
\begin{array}{c}
\mbox{---}
\\
\ket{0}\bra{0}
\end{array}
\;.
\eeq
Then the Szegedy quantum
walk operator $W$ for the
Bayesian net $\rvx$
is defined
by

\beq
W = U(-1)^{\check{\pi}}U^\dagger (-1)^{\hat{\pi}}
\;.
\eeq
\subsection{Eigenvalues of $W$}

To find the eigenvalues of $W$,
we will use the following identities.

\begin{claim}\label{cl-proj-ids}
\begin{subequations}
\beq
\hat{\pi}\ket{m_j 0} = \ket{m_j 0}
\;,
\eeq

\beq
\hat{\pi}(U\swap)\ket{m_j 0} = m_j\ket{m_j 0}
\;,\label{eq-pi-u-swap}
\eeq

\beq
\hat{\pi}(\swap U^\dagger)\ket{m_j 0} = m^*_j\ket{m_j 0}
\;,\label{eq-pi-swap-u}
\eeq
\end{subequations}
for all $j\in \{0,1,\ldots,N_S-1\}$.
\end{claim}
\proof

From the definition of $\hat{\pi}$,
we see that
\beq
\hat{\pi}
\begin{array}{l}
\ket{0}
\\
\ket{m_j}
\end{array}
=
\begin{array}{l}
\ket{0}
\\
\ket{m_j}
\end{array}
\;.
\eeq
Also,

\beq
\hat{\pi}
(U\swap)
\begin{array}{l}
\ket{0}
\\
\ket{m_j}
\end{array}
=
\sum_k
\begin{array}{r}
\ket{0}\bra{0}
\\
\ket{m_k}\bra{m_k}
\end{array}
U
\begin{array}{l}
\ket{m_j}
\\
\ket{0}
\end{array}
=
m_j
\begin{array}{l}
\ket{0}
\\
\ket{m_j}
\end{array}
\;,
\eeq
and

\beq
\hat{\pi}
(\swap U^\dagger)
\begin{array}{l}
\ket{0}
\\
\ket{m_j}
\end{array}
=
\sum_k
\begin{array}{r}
\ket{0}\bra{m_k}
\\
\ket{m_k}\bra{0}
\end{array}
U^\dagger
\begin{array}{l}
\ket{0}
\\
\ket{m_j}
\end{array}
=
m^*_j
\begin{array}{l}
\ket{0}
\\
\ket{m_j}
\end{array}
\;.
\eeq
\qed

An immediate
consequence of Claim \ref{cl-proj-ids} is that

\beq
\bra{m_{j} 0}
U\swap \ket{m_k 0}
=
\bra{m_{j} 0}\hat{\pi}
U\swap \ket{m_k 0}
=
m_j\delta_j^{k}
\;,
\label{eq-u-swap-mat-elem}
\eeq
for $j,k\in\{0,1,\ldots,\ns-1\}$.

Note that since $m_0=1$,
Eq.(\ref{eq-u-swap-mat-elem}) implies that

\beq
\ket{m_00} = U\swap\ket{m_00}
\;.
\label{eq-u-swap-equals-1}
\eeq

Another consequence
of Claim \ref{cl-proj-ids}
is that $\ket{m_0 0}$
is a stationary state of $W$. Indeed,
one has

\beqa
W\ket{m_0 0}&=&
U
(-1)^{\check{\pi}}
U^\dagger
(-1)^{\hat{\pi}}
\ket{m_0 0}
\\
&=&
U
\swap
(1-2\hat{\pi})
\swap
U^\dagger
(-1)
\ket{m_0 0}
\\
&=&
(1-2m_0U\swap)(-1)
\ket{m_0 0}
\\
&=&
(1-2)(-1)
\ket{m_0 0}
\\
&=&\ket{m_0 0}
\;.
\eeqa

Let
\beq
\calv_{busy}^j=span\{\ket{m_j0}, U\swap\ket{m_j0}\}
\;
\eeq
for $j\in\{0,1,\ldots,N_S-1\}$.
(By ``span" we mean all
linear combinations of
these vectors with {\it complex}
coefficients.)

\begin{claim}\label{cl-v-busy-j}
$W\calv_{busy}^j=\calv_{busy}^j$
for all $j\in\{0,1,\ldots,N_S-1\}$.
For $j=0$, let

\beq
\ket{\psi_0} = \ket{m_0 0}
\;.
\eeq
$\{\ket{\psi_0}\}$ is an
orthonormal basis for $\calv^0_{busy}$
and $W\ket{\psi_0} = \ket{\psi_0}$.
For $j\neq 0$, let

\beq
\ket{\psi_{\pm j}}=
\frac{\pm i}{\sqrt{2} \sin\varphi_j}
(e^{-i\eta_j}U\swap\ket{m_j0}- e^{\pm i2\varphi_j}\ket{m_j0})
\;.
\label{eq-psi-j-original-basis}
\eeq
$\{\ket{\psi_j}, \ket{\psi_{-j}}\}$ is an
orthonormal basis for $\calv^j_{busy}$
and $W\ket{\psi_{\pm j}} =
e^{\pm i2\varphi_j}\ket{\psi_{\pm j}}$.
\end{claim}
\proof

Using the identities of
Claim \ref{cl-proj-ids},
one finds after some algebra that

\begin{subequations}
\label{eq-w-invariant-plane}
\beq
W\ket{m_j 0}=
(-1)\ket{m_j 0}
+
(2m^*_j)U\swap \ket{m_j 0}
\;,
\eeq
and

\beq
W(U\swap)\ket{m_j 0}=
(-2m_j)\ket{m_j 0}
+
(-1 + 4|m_j|^2)U\swap \ket{m_j 0}
\;
\eeq
\end{subequations}
for all $j$.

According
to Eqs.(\ref{eq-w-invariant-plane}),
$\calv_{busy}^j$ is invariant under
the action of $W$ for each $j$.
By virtue of Eq.(\ref{eq-u-swap-mat-elem}),
$\calv_{busy}^j$
is 1-dim for $j=0$ and 2-dim if $j\neq 0$.
We've already proven that
$\ket{m_0 0}$
is a stationary state of $W$.

\begin{figure}[h]
    \begin{center}
    \epsfig{file=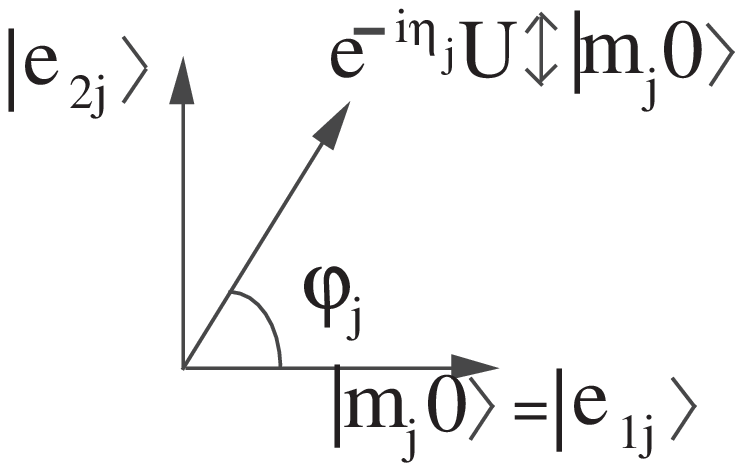, height=1in}
    \caption{Definition of $\ket{e_{1j}}$
    and $\ket{e_{2j}}$.
    }
    \label{fig-e1-e2-def}
    \end{center}
\end{figure}

Now consider a fixed $j\neq0$.
Both $U(-1)^{\check{\pi}}U^\dagger$ and
$(-1)^{\hat{\pi}}$ are
reflections,
and reflections are a special type
of orthogonal matrix, so the product
of these 2 orthogonal matrices is
also an orthogonal matrix.
In fact, it's a rotation
about the axis perpendicular
to the planar subspace $\calv_{busy}^j$.
The vectors $\ket{m_j0}$,
and $U\swap\ket{m_j0}\}$ are
independent but not orthogonal.
However, we can express them in terms of orthogonal vectors
(see Fig.\ref{fig-e1-e2-def}) as follows:

\begin{subequations}
\beq
\ket{m_j0} = \ket{e_{1j}}
\;,
\eeq
and

\beq
e^{-i\eta_j}U\swap\ket{m_j0} =
\cos(\varphi_j)\ket{e_{1j}}+
\sin(\varphi_j)\ket{e_{2j}}
\;.
\eeq
\end{subequations}
In the $\ket{e_{1j}}$, $\ket{e_{2j}}$ basis,
we find after substituting
$m_j = e^{i\eta_j}\cos(\varphi_j)$ into
Eqs.(\ref{eq-w-invariant-plane}) that

\beq
W=
\left[
\begin{array}{cc}
\cos(2\varphi_j)
&\sin(2\varphi_j)\\
-\sin(2\varphi_j)
&\cos(2\varphi_j)
\end{array}
\right]
\;.
\eeq
The eigenvalues of this matrix
are $e^{\pm i2\varphi_j}$, with
corresponding eigenvectors:

\beq
\ket{\psi_{\pm j}} =
\frac{1}{\sqrt{2}}(\ket{e_{1j}}\pm \ket{e_{2j}})
\;.
\label{eq-psi-j-e-basis}
\eeq
These eigenvectors satisfy

\beq
\av{\psi_{-j}|\psi_{j}}=0\;\;,\;\;
\av{\psi_{j}|\psi_{j}}
= \av{\psi_{-j}|\psi_{-j}}=1
\;.
\eeq
By expressing $\ket{e_{1j}}$
and $\ket{e_{2j}}$ in
Eq.(\ref{eq-psi-j-e-basis})
in the original basis,
we get Eq.(\ref{eq-psi-j-original-basis}).
\qed

Define the following vector spaces:

\beq
\calv= span\{ \ket{x}\otimes\ket{y}: x,y\in S_\rvx\}
\;,
\eeq

\beq
\calv_A= span\{ \ket{x}\otimes\ket{0}: x\in S_\rvx\}
\;,
\eeq

\beq
\calv_B = U\swap \calv_A
\;,
\eeq
and

\beq
\calv_{busy} = \calv_A + \calv_B
\;.
\eeq
$\calv$ can be expressed as
a direct sum of
$\calv_{busy}$ and its orthogonal
complement $\calv_{busy}^\perp$:

\beq
\calv=\calv_{busy}\oplus \calv_{busy}^\perp
\;.
\eeq
From Claim \ref{cl-v-busy-j},
it follows that
$\calv_{busy}$
is a direct sum of the subspaces $\calv_{busy}^j$:

\beq
\calv_{busy} =\bigoplus_{j=0}^{N_S-1} \calv_{busy}^j
\;.
\label{eq-v-busy-direct-sum}
\eeq
Recall that matrices
$M_1,M_2$ and $M_{hyb}$ are $N_S$ dimensional
whereas $W$ is $N_S^2$ dimensional.
Since
the size of $S_\rvx$
is $N_S$,
$dim(\calv) = N_S^2$.
From Eq.(\ref{eq-v-busy-direct-sum}) and
Claim \ref{cl-v-busy-j},
$dim(\calv_{busy})=2N_S-1$.
Furthermore,
$\{\ket{\psi_j}: j=0, \pm 1, \pm 2,
\ldots, \pm(\ns-1)\}$
is an orthonormal basis for $\calv_{busy}$.

At this point we've
explained the action
of $W$ on $\calv_{busy}$,
but we haven't said anything
about the action of $W$ on
$\calv^\perp_{busy}$. Next we
show that $W$
acts simply as the identity
on $\calv^\perp_{busy}$.
(This is what one would expect since
the vectors in
$\calv^\perp_{busy}$ are parallel
to the axis of the $W$ rotation.)
Recall that if $S$ and $T$
are subspaces of a vector space $\calv$,
then
$(S+T)^\perp = S^\perp\cap T^\perp$.
Therefore,

\beq
\calv^\perp_{busy} = \calv_A^\perp\cap \calv_B^\perp
\;.
\eeq
From the definitions
of $\calv_A$ and $\calv_B$,
it's easy to see that

\beq
\calv_A^\perp=span\{\ket{x}\otimes\ket{y}:
x\in S_\rvx,
\mbox{ and } y\in S_\rvx-\{0\}\}
\;,
\eeq
and

\beq
\calv_B^\perp=U\swap(\calv_A^\perp)
\;.
\eeq

\begin{claim}
\beq
W\ket{\phi} = \ket{\phi}
\;
\eeq
for all $\ket{\phi}\in \calv_{busy}^\perp$.
\end{claim}
\proof
Let $\ket{\phi}\in \calv_{busy}^\perp=
\calv_A^\perp \cap \calv_B^\perp$. Hence
$\ket{\phi}\in \calv_A^\perp$
and $\ket{\phi}=U\swap \ket{\theta}$
for some $\ket{\theta}\in \calv_A^\perp$.

\beqa
U
(-1)^{\check{\pi}}
U^\dagger
(-1)^{\hat{\pi}}
\ket{\phi}
&=&
U
\swap
(-1)^{\hat{\pi}}
\swap
U^\dagger(-1)^0
\ket{\phi}
\\
&=&
U
\swap
(1-2\hat{\pi})
\swap
U^\dagger
U\swap\ket{\theta}
\\
&=&
U
\swap
(1-2\hat{\pi})
\ket{\theta}
\\
&=&
\ket{\phi}
\;.
\eeqa
\qed

It's interesting to
compare
the present paper
with Ref.\cite{TucQusann}.
For  Ref.\cite{TucQusann},
$M_1=M_2=M$
and $\pi()$
 is a standard detailed balance
for $M$
instead of a detailed balance for the pair $(M_1,M_2)$.
For  Ref.\cite{TucQusann}, $M_{hyb}=M_{sym}$,
$m_j=m^*_j$,
$U_1=\check{U}$,
$U_2=\hat{U}$,
$U=U_2^\dagger U_1 = \hat{U}^\dagger \check{U}$,
$U=\swap U^\dagger \swap$.
When $U=\swap U^\dagger \swap$
as in Ref.\cite{TucQusann},
Eq.(\ref{eq-pi-u-swap}) and
Eq.(\ref{eq-pi-swap-u})
are essentially identical,
whereas in the $M_1\neq M_2$
case, it's less obvious
that
these two equations are true simultaneously.

\section{Quantum Gibbs Sampling Algorithm}
In this section,
we will describe an
algorithm for doing Gibbs sampling
on a quantum computer,
utilizing the Szegedy
operator $W$ that we
have so painstakingly discussed
in previous sections.

We begin by
choosing\footnote{Perhaps some symmetry of
the physical situation being
modeled by the Bayesian network $\rvx$
will suggest some $\rvx$
value that has non-zero probability.
Alternatively, one can proceed as follows.
For definiteness, consider
a Bayesian net $\rvx=(\rvx_1,\rvx_2,\rvx_3)$
with 3 nodes. Suppose
$P(x_3,x_2,x_1)=P(x_3|x_2,x_1)P(x_2|x_1)P(x_1)$
and the functions
$P_{\rvx_3|\rvx_2,\rvx_1}$
$P_{\rvx_2|\rvx_1}$
and $P_{\rvx_1}$ are known.
Choose $y_1\in S_{\rvx_1}$ such that
$P_{\rvx_1}(y_1)\neq 0$.
Then choose $y_2\in S_{\rvx_2}$ such that
$P_{\rvx_2|\rvx_1}(y_2|y_1)\neq 0$.
Finally, choose $y_3\in S_{\rvx_3}$ such that
$P_{\rvx_3|\rvx_2,\rvx_1}(y_3|y_2,y_1)\neq 0$.
Set $x_0=(y_1,y_2,y_3)$.}
some $x_0\in S_\rvx$
for which $P(\rvx=x_0)\neq 0$.
Now define

\beq
\ket{x_0 0}=
\ket{\rvx=x_0}\otimes\ket{0}^{\otimes \nb}
\;.
\eeq
Note that $\ket{x_0 0}\in \calv_{busy}$
and

\beq
\av{\psi_0|x_0 0}=
\av{\sqrt{\pi}|\rvx=x_0}=\sqrt{\pi(x_0)}
\;.
\eeq
$\sqrt{\pi(x)}=\sqrt{P(x)}$
can be easily evaluated at a single
point $x=x_0$.
Our quantum
Gibbs algorithm
consists of performing
the original Grover algorithm
with beginning state $\ket{x_00}$
and target state $\ket{\psi_0}$.
Define the following
2 reflection operators

\beq
R_{beg}=(-1)^{\ket{x_0 0}\bra{x_0 0}}
\;,
\eeq
and

\beq
R_{tar} = (-1)^{\ket{\psi_0}\bra{\psi_0}}
\;.
\eeq
$R_{beg}R_{tar}$ is a rotation
by an angle $\calo(\sqrt{\pi(x_0)})$
in space $span\{\ket{\psi_0}, \ket{x_00}\}
\subset \calv_{busy}$.
Let

\beq
L=\calo(\frac{1}{\sqrt{\pi(x_0)}})
\;.
\eeq
If $\sqrt{\pi(x_0)}=\calo(1/\sqrt{\ns})$,
then $L$ iterations
of $R_{beg}R_{tar}$ will take the beginning
state to the target state.\footnote{We
will discuss in a future paper
what to do if
$\sqrt{\pi(x_0)}$ is much
larger than $\calo(1/\sqrt{\ns})$.}
To
implement this
use of Grover's algorithm,
we need to compile
(with polynomial efficiency)
the operator $R_{beg}R_{tar}$.
$R_{beg}$
is easy to compile; it's just
a single multiply-controlled phase.
Next, we will
explain how to compile $R_{tar}$.

\begin{figure}[h]
    \begin{center}
    \epsfig{file=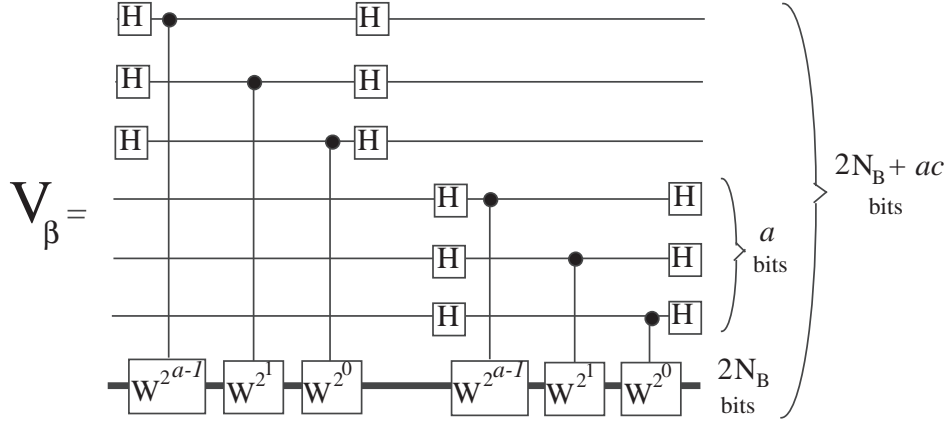, height=2.25in}
    \caption{Definition of operator
    $V_\beta$ for Szegedy operator $W$.}
    \label{fig-v-beta}
    \end{center}
\end{figure}

Fig.\ref{fig-v-beta},
which is identical
to Fig.18 in Ref.\cite{TucQusann},
defines an operator $V_\beta$ in
terms of multiple (modified)
phase estimation steps.
The $V_\beta$ of Ref.\cite{TucQusann}
depends on a parameter
$\beta$ (inverse temperature)
because the operator $M$ in that paper
depends on this parameter.
$V_\beta$ in the present paper does not depend
on  $\beta$
(because the Bayesian net $\rvx$ doesn't)
so we will drop the $\beta$
subscript from it henceforth, and refer
to it simply as $V$.
$V$ does not depend on $\beta$ but
it still depends on the positive
integers $a$ and $c$.
(In the language of Ref.\cite{TucQusann},
$a$=number of probe bits for
each PE (Phase Estimation) step, and
$c$=number of PE
steps). Note that operator $W$ is applied
$2^a c$ times by $V$.

Let $\ket{0^{ac}}=\ket{0}^{\otimes(ac)}$,
$J=\{0,\pm 1,\pm 2, \ldots, \pm (\ns-1)\}$, and
 $J'=J-\{0\}$.
According to Lemma 2 of Ref.\cite{wocjan1},
for any $\epsilon_2>0$, if
we adjust the integers
$a$ and $c$ so that

\beq
2^a \approx \frac{1}{\Delta}=\calo(\frac{1}{\sqrt{\delta}})
\;,
\eeq
and

\beq
c\approx \log_2(\frac{1}{\sqrt{\epsilon_2}})
\;,
\eeq
(recall $\delta=1-|m_1|$ is the distance between the
two largest
eigenvalue magnitudes of $M_1$),
then
$V$ acts
on
the space $\calv_{busy}\otimes \ket{0^{ac}}$
as follows:

\beq
V=
\begin{array}{c}
\ket{0^{ac}}\bra{0^{ac}}\\
\ket{\psi_0}\bra{\psi_0}
\end{array}
+
\sum_{j\in J'}
\begin{array}{c}
\ket{\chi_j}\bra{0^{ac}}
\\
\ket{\psi_j}\bra{\psi_j}
\end{array}
+
\calo(\sqrt{\epsilon_2})
\;,
\label{eq-v-approx}
\eeq
where the
$\ket{\chi_j}$ are states
of $ac$ qubits
such that $\av{0^{ac}|\chi_j}=0$.
In Eq.(\ref{eq-v-approx})
and for the remainder
of this section, the
top row represents the $ac$ ancilla qubits
shown in Fig.\ref{fig-v-beta},
and the bottom row represents the
$2\nb$ qubits on which $W$ operates.

Now define
\beq
Q= (-1)^{\ket{0^{ac}}\bra{0^{ac}}}
= 1 - 2\ket{0^{ac}}\bra{0^{ac}}
\;,
\eeq
and

\beq
\tilde{R}_{tar}=
V^\dagger
\begin{array}{c}
Q\\
\mbox{---}
\end{array}
V
\;.
\eeq
It follows that for
any $\ket{\psi}\in\calv_{busy}$,

\beqa
\tilde{R}_{tar}
\begin{array}{l}
\ket{0^{ac}}
\\
\ket{\psi}
\end{array}
&=&
\left[
1-
2V^\dagger
\begin{array}{c}
\ket{0^{ac}}\bra{0^{ac}}\\
\mbox{---}
\end{array}
V
\right]
\begin{array}{l}
\ket{0^{ac}}
\\
\ket{\psi}
\end{array}
\\
&=&
\left[
1-
2
\begin{array}{c}
\ket{0^{ac}}\bra{0^{ac}}\\
\ket{\psi_0}\bra{\psi_0}
\end{array}
\right]
\begin{array}{l}
\ket{0^{ac}}
\\
\ket{\psi}
\end{array}
+
\calo(\sqrt{\epsilon_2})
\\
&=&
\begin{array}{l}
\ket{0^{ac}}
\\
\ket{\psi}
\end{array}
+
\begin{array}{l}
\ket{0^{ac}}
\\
(-2\ket{\psi_0}\bra{\psi_0})\ket{\psi}
\end{array}
+ \calo(\sqrt{\epsilon_2})
\\
&=&
\begin{array}{l}
\ket{0^{ac}}
\\
R_{tar}\ket{\psi}
\end{array}
+ \calo(\sqrt{\epsilon_2})
\;.
\label{eq-r-tilde-tar-and-r-tar}
\eeqa
Eq.(\ref{eq-r-tilde-tar-and-r-tar})
is the essence of Corollary 2 in Ref.\cite{wocjan1}.
It means that
$R_{tar}$ acting on $\calv_{busy}$
 can be approximated
by $\tilde{R}_{tar}$ acting on
$\calv_{busy}\otimes \ket{0^{ac}}$.
Since we already know
how to compile $\tilde{R}_{tar}$,
we have accomplished
our goal of compiling $R_{tar}$,
at least approximately.

Next, we will try
to estimate
the error of our
quantum Gibbs algorithm.
Suppose $\tilde{\pi}()$
is our estimate of $\pi()$.
Note that for any $x\in S_\rvx$,

\beqa
|\pi(x)-\tilde{\pi}(x)|
&=&
|(\sqrt{\pi(x)}-\sqrt{\tilde{\pi}(x)})
(\sqrt{\pi(x)}+\sqrt{\tilde{\pi}(x)})|
\\
&\leq&
2|\sqrt{\pi(x)}-\sqrt{\tilde{\pi}(x)}|
\;.
\eeqa
Suppose $\epsilon>0$ is defined so that

\beq
\max_x|\sqrt{\pi(x)}-\sqrt{\tilde{\pi}(x)}|\leq \epsilon
\;.
\eeq
Then, since we apply
the  $R_{beg}R_{tar}$ operator
a total of $L$ times,
and each time we can incur
an error of $\sqrt{\epsilon_2}$,

\beq
\epsilon\approx L \sqrt{\epsilon_2}
\;.
\eeq
If we define one
step as one $W$ application,
then the total number of steps
for the whole algorithm is
$\calo(L2^a c)=
\calo(\frac{L}{\sqrt{\delta}}
\log_2(\frac{L}{\epsilon}))$.
Thus, our algorithm will
yield a sample of the classical
Bayesian net $\rvx$ with precision
$\calo(\epsilon)$, in
$\calo(\frac{L}{\sqrt{\delta}}
\log_2(\frac{L}{\epsilon}))$
steps. Achieving
the same precision with
a classical Gibbs sampling
algorithm would require
${\cal O}(\frac{1}{\delta})$
steps.

The Szegedy
operator $W$
of this paper
can also be used
to do quantum
simulated annealing
and Metropolis-Hastings
if
the marginals
$P(x^{t+1}_i|x^{t}_\noti)$
can be calculated for
each $i$
from the transition
matrix $P(x^{t+1}|x^{t})$.
(In the case of simulated
annealing, $P(x^{t+1}|x^{t})$
is different for each $\beta_i$
of the annealing schedule).

\end{document}